\documentclass[prl]{revtex4}

\usepackage{graphicx}
\usepackage{amssymb}
\usepackage{amsmath}
\usepackage{amsfonts}
\usepackage{epstopdf}
\usepackage{epsfig}
\usepackage{wrapfig}

\usepackage{bm}

\newcommand{\beq}{\begin{equation}}
\newcommand{\eeq}{\end{equation}}
\newcommand{\beqar}{\begin{eqnarray}}
\newcommand{\eeqar}{\end{eqnarray}}
\newcommand{\bea}{\begin{eqnarray}}
\newcommand{\eea}{\end{eqnarray}}
\newcommand{\bcen}{\begin{center}}
\newcommand{\ecen}{\end{center}}

\begin{document}

\title{Comments on "cooling by heating: Refrigeration powered by Photons'' } 
\author{Amikam Levy$^{(1)}$, Robert Alicki$^{(2)(3)}$ and Ronnie Kosloff$^{(1)}$ }
\affiliation{$^{(1)}$Institute  of Chemistry  The Hebrew University, Jerusalem 91904, Israel}
\affiliation{$^{(2)}$Institute of Theoretical Physics and Astrophysics, University of
Gda\'nsk, Poland }
\affiliation{$^{(3)}$Weston Visiting Professor, Weizmann Institute of Science, Rehovot, Israel}

\maketitle

In a recent letter, Cleuren et. al. \cite{cleuren12} proposed a mechanism for solar refrigeration composed of two metallic leads  
mediated by two coupled quantum dots and powered by (solar) photons. In their analysis the refrigerator can operate to
$T_r \rightarrow 0$ and the cooling flux $\dot Q_r \propto T_r$.  We comment that this model violates the dynamical version of the III-law of thermodynamics.
\par
There are seemingly two independent formulation of the third law. 
The first, known as the Nernst heat theorem,  implies that the entropy flow from any substance at absolute zero temperature is zero.
At steady state the second law implies that the total entropy production is non-negative, $\sum_i-\frac{\dot Q_i}{T_i} \geq 0$ where $\dot Q_i$ is positive for heat flowing into the system from the $i$-th bath.
In order to insure the fulfillment of the second law when one of the heat baths (labeled $k$) approaches the absolute zero temperature. 
It is necessary that the entropy production from this bath scales as $\dot S_k \sim T_k^{\alpha}$ with $\alpha \geq 0$. 
For the case where $\alpha=0$ the fulfillment of the second law depends on the entropy production of the other baths, which should compensate on the negative entropy production of the $k$ bath. 
The first formulation of the third law slightly modifies this restriction. Instead of $\alpha \geq 0$ the third low impose $\alpha > 0 $ guaranteeing that at the absolute zero $\dot S_ k = 0$. 
\par
The second formulation of the third law  is a dynamical one, known as the unatinability principle:  \emph{No refrigerator can cool a system to absolute zero temperature at finite time}.
This formulation is more restrictive, imposing limitations on the spectral density and the dispersion dynamics of the heat bath \cite{levy12}. 
We quantify this formulation by evaluating the characteristic exponent $\zeta$ of the cooling process
\begin{equation}
\label{eq:cp}
 \frac{dT(t)}{dt} \sim -T^{\zeta}, ~~~ T\rightarrow 0 
\end{equation}
 Namely for $\zeta < 1$ the system is cooled to zero temperature at finite time. 
Eq.(\ref{eq:cp}) can be related to the heat flow:
\begin{equation}
 \dot Q_k(T_k(t)) = -c_V(T_k(t))\frac{dT_k(t)}{dt}
\label {eq:3law}
\end{equation}
where $c_V$ is the heat capacity of the  bath. 
\par
The refrigerator presented in \cite{cleuren12} violates the III-law as in Eqs. (\ref{eq:3law}) and (\ref{eq:cp}).
For an electron reservoir at low temperature the heat capacity  $c_V \sim T$. The heat current of the refrigerator of \cite{cleuren12} 
$\dot Q_r \propto T_r$ therefore one obtains $\zeta=0$ hence zero temperature is achieved at finite time, in contradiction with the third law. 
\par
Finding the flow in the analysis of  \cite{cleuren12} is not a trivial task. A possible explanation emerges from the assumption in \cite{cleuren12}
that transitions between lower and higher levels within the individual dots are negligible. 
Photon assisted tunneling between dots produce a week tunnel current \cite{vanderwiel03}. In comparison quenching transitions in the individual dots cannot be neglected. A modified master equation which includes these transitions can be constructed
for  a five level system:  $ \dot {\vec{p}} =M \cdot \vec{p}$  where $\vec p =(p_0, p_{ld},p_{rd},p_{lu},p_{ru})^{T} $.
Where $p_0$ is the probability of finding no electron in the double dot and $p_{ij}$ is the probability of finding one electron in the corresponding energy level,
with $l$-left, $r$-right, $d$-down, $u$-up.
The $M$ matrix is 5x5 matrix which includes also quenching transition in the individual dots (see appendix).
\par
A crucial condition for the device to operate as a refrigerator \cite{cleuren12} is that there is no net electric charging of the baths.
Otherwise the electric current flowing through the device must be compensated by an external flow of electrons from the hot to the cold bath which  would annihilate the cooling effect.
When assuming that the relaxation rates within the individual dots are equal and finite, one obtains that the two manifolds of cooling and no net charging do not intersect in parameter space.
As a result the condition for cooling and no net charging in the baths cannot be satisfied simultaneously.
\par
In conclusion, transitions in the individual dots, which are always present in real system, can not be neglected when treating electron transport in the double dot.
The fulfillment of thermodynamical laws are a strong tool for verifying the quantum description of such nano-devices.
For quantum description of refrigerator powered by heat (absorption refrigerator) see \cite{levy12,levy212}.  

\section{Appendix} 
The $M$ matrix reads
\small{
\[
M =
\left( {\begin{array}{ccccc}
 -(k_{b\rightarrow s}^{\epsilon_1-\epsilon_g}  +k_{b\rightarrow s}^{\epsilon_1}+k_{b\rightarrow s}^{\epsilon_2+\epsilon_g}  +k_{b\rightarrow s}^{\epsilon_2})
 & k_{s\rightarrow b}^{\epsilon_1-\epsilon_g} & k_{s\rightarrow b}^{\epsilon_1} & k_{s\rightarrow b}^{\epsilon_2+\epsilon_g} & k_{s\rightarrow b}^{\epsilon_2}  \\ \\ 
k_{b\rightarrow s}^{\epsilon_1-\epsilon_g} & -k_{s\rightarrow b}^{\epsilon_1-\epsilon_g} -k_{\uparrow}^{\epsilon_g}-k_{\uparrow}^{\Delta_l} & k_{\downarrow}^{\epsilon_g}& k_{\downarrow}^{\Delta_l} & 0 \\ \\ 
k_{b\rightarrow s}^{\epsilon_1} & k_{\uparrow}^{\epsilon_g} & -k_{s\rightarrow b}^{\epsilon_1}-k_{\downarrow}^{\epsilon_g}-k_{\uparrow}^{\Delta_r} & 0 & k_{\downarrow}^{\Delta_r} \\ \\
k_{b\rightarrow s}^{\epsilon_2+\epsilon_g} & k_{\uparrow}^{\Delta_l} & 0 & -k_{s\rightarrow b}^{\epsilon_2+\epsilon_g} -k_{\downarrow}^{\epsilon_g}-k_{\downarrow}^{\Delta_l} & k_{\uparrow}^{\epsilon_g} \\  \\
k_{b\rightarrow s}^{\epsilon_2} & 0 & k_{\uparrow}^{\Delta_r} & k_{\downarrow}^{\epsilon_g} & -k_{s\rightarrow b}^{\epsilon_2} -k_{\uparrow}^{\epsilon_g}-k_{\downarrow}^{\Delta_r} 
\end{array} } \right)
\]} 
The energy difference between the lower and upper levels in the left and right dots are $\Delta_l = \epsilon_2 -\epsilon_1 +2 \epsilon_g$ and $\Delta_r = \epsilon_2-\epsilon_1$ respectively.
The transition rate between the bath and system is given by:
\begin{eqnarray}
\begin{array}{c}
k_{b\rightarrow s}^{\epsilon}= \Gamma f(\epsilon)\\ 
k_{s\rightarrow b}^{\epsilon}= \Gamma (1- f(\epsilon))\\ 
f(\epsilon)=[exp((\epsilon -\mu)/T)+1]^{-1} 
\end{array}
\end{eqnarray}
The transition rates between dots are given by:
\begin{eqnarray}
\begin{array}{c}
k_{\uparrow}^{\epsilon}= \Gamma_s n(\epsilon)\\ 
k_{\downarrow}^{\epsilon}= \Gamma_s (1+ n(\epsilon))\\ 
n(\epsilon)=[exp(\epsilon/T_s)-1]^{-1} 
\end{array}
\end{eqnarray}

\begin{figure}[htbp]
\center{\includegraphics[height=3cm,width=8cm]{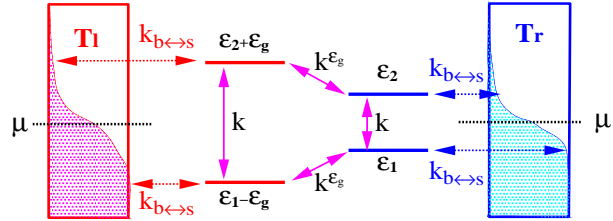}}
\caption{Schematic representation of all possible electron transitions}
\label{fig:1}
\end{figure}

In the high temperature limit $k_{\downarrow}^{\epsilon_g} \simeq k_{\uparrow}^{\epsilon_g}$. 
For simplicity we assume that $k_{\downarrow}^{\Delta_r}=k_{\uparrow}^{\Delta_r}=k_{\downarrow}^{\Delta_l}=k_{\uparrow}^{\Delta_l}\equiv k$, see Fig \ref{fig:1} .  
We define $J_{rd} (J_{ru})$ as the particle current from the right bath into the $\epsilon_1 (\epsilon_2)$ level and $J_{ld} (J_{lu})$ as the particle current from the left bath into the $\epsilon_1 -\epsilon_g (\epsilon_2 + \epsilon_g)$ level, respectively.
\begin{eqnarray}
\begin{array}{c}
J_{rd}= k_{b\rightarrow s}^{\epsilon_1} p_0 - k_{s\rightarrow b}^{\epsilon_1} p_{rd} \\  \\
J_{ru}= k_{b\rightarrow s}^{\epsilon_2} p_0 - k_{s\rightarrow b}^{\epsilon_2} p_{ru} \\  \\
J_{ld}= k_{b\rightarrow s}^{\epsilon_1 -\epsilon_g} p_0 - k_{s\rightarrow b}^{\epsilon_1-\epsilon_g} p_{ld} \\  \\
J_{lu}= k_{b\rightarrow s}^{\epsilon_2 +\epsilon_g} p_0 - k_{s\rightarrow b}^{\epsilon_2+\epsilon_g} p_{lu}  
\end{array}
\end{eqnarray} 

Negative current imply transition of particle from the system into the bath. The heat currents are given by:
\begin{eqnarray}
\begin{array}{c}
\dot{Q}_r =(\epsilon_1 -\mu)J_{rd} + (\epsilon_2-\mu)J_{ru}\\ 
\dot{Q}_l =(\epsilon_1 -\epsilon_g -\mu)J_{ld} + (\epsilon_2 +\epsilon_g -\mu)J_{lu}
\end{array}
\end{eqnarray}
As in \cite{cleuren12} we take $\epsilon_2=-\epsilon_1=\epsilon$ and $\mu=0$.
The condition of no net charging is $J_{rd}+J_{ru}=J_{ld}+J_{lu}=0$ which eventually reduces to $\frac{\epsilon}{T_r}=\frac{\epsilon+\epsilon_g}{T_l}$ (note that for $\epsilon_g=0$ no net charging is translated to $T_l=T_r$).  
Calculating the heat flows under this condition leads to:
\begin{eqnarray}
\begin{array}{c}
\dot{Q}_r =-\dfrac{4k\Gamma \epsilon sinh(\frac{\epsilon+\epsilon_g}{T_l})}{10k+\Gamma+(10k+4\Gamma)cosh(\frac{\epsilon+\epsilon_g}{T_l})} \\ 
\dot{Q}_l =-\dfrac{4k\Gamma (\epsilon+\epsilon_g) sinh(\frac{\epsilon+\epsilon_g}{T_l})}{10k+\Gamma+(10k+4\Gamma)cosh(\frac{\epsilon+\epsilon_g}{T_l})}
\end{array}
\end{eqnarray}
Both currents are negative implying that heat will always flow from the photon source into the baths.
Therefor no refrigeration occurs.     
\bibliography{citeamikam}

\begin{thebibliography}{4}
\expandafter\ifx\csname natexlab\endcsname\relax\def\natexlab#1{#1}\fi
\expandafter\ifx\csname bibnamefont\endcsname\relax
  \def\bibnamefont#1{#1}\fi
\expandafter\ifx\csname bibfnamefont\endcsname\relax
  \def\bibfnamefont#1{#1}\fi
\expandafter\ifx\csname citenamefont\endcsname\relax
  \def\citenamefont#1{#1}\fi
\expandafter\ifx\csname url\endcsname\relax
  \def\url#1{\texttt{#1}}\fi
\expandafter\ifx\csname urlprefix\endcsname\relax\def\urlprefix{URL }\fi
\providecommand{\bibinfo}[2]{#2}
\providecommand{\eprint}[2][]{\url{#2}}

\bibitem[{\citenamefont{Cleuren et~al.}(2012)\citenamefont{Cleuren, Rutten, and
  Van~den Broeck}}]{cleuren12}
\bibinfo{author}{\bibfnamefont{B.}~\bibnamefont{Cleuren}},
  \bibinfo{author}{\bibfnamefont{B.}~\bibnamefont{Rutten}}, \bibnamefont{and}
  \bibinfo{author}{\bibfnamefont{C.}~\bibnamefont{Van~den Broeck}},
  \bibinfo{journal}{Phys. Rev. Lett.} \textbf{\bibinfo{volume}{108}},
  \bibinfo{pages}{120603} (\bibinfo{year}{2012}).

\bibitem[{\citenamefont{Levy and Kosloff}(2012)}]{levy12}
\bibinfo{author}{\bibfnamefont{A.}~\bibnamefont{Levy}} \bibnamefont{and}
  \bibinfo{author}{\bibfnamefont{R.}~\bibnamefont{Kosloff}},
  \bibinfo{journal}{Phys. Rev. Lett.} \textbf{\bibinfo{volume}{108}},
  \bibinfo{pages}{070604} (\bibinfo{year}{2012}).

\bibitem[{\citenamefont{van~der Wiel et~al.}(2002)\citenamefont{van~der Wiel,
  De~Franceschi, Elzerman, Fujisawa, Tarucha, and Kouwenhoven}}]{vanderwiel03}
\bibinfo{author}{\bibfnamefont{W.~G.} \bibnamefont{van~der Wiel}},
  \bibinfo{author}{\bibfnamefont{S.}~\bibnamefont{De~Franceschi}},
  \bibinfo{author}{\bibfnamefont{J.~M.} \bibnamefont{Elzerman}},
  \bibinfo{author}{\bibfnamefont{T.}~\bibnamefont{Fujisawa}},
  \bibinfo{author}{\bibfnamefont{S.}~\bibnamefont{Tarucha}}, \bibnamefont{and}
  \bibinfo{author}{\bibfnamefont{L.~P.} \bibnamefont{Kouwenhoven}},
  \bibinfo{journal}{Rev. Mod. Phys.} \textbf{\bibinfo{volume}{75}},
  \bibinfo{pages}{1} (\bibinfo{year}{2002}).

\bibitem[{\citenamefont{Levy et~al.}(2012)\citenamefont{Levy, Alicki, and
  Kosloff}}]{levy212}
\bibinfo{author}{\bibfnamefont{A.}~\bibnamefont{Levy}},
  \bibinfo{author}{\bibfnamefont{R.}~\bibnamefont{Alicki}}, \bibnamefont{and}
  \bibinfo{author}{\bibfnamefont{R.}~\bibnamefont{Kosloff}},
  \bibinfo{journal}{Phys. Rev. E} \textbf{\bibinfo{volume}{85}},
  \bibinfo{pages}{061126} (\bibinfo{year}{2012}).

\end{thebibliography}

\end{document}